\documentclass[11pt]{article} 
\usepackage{amsfonts,amscd,amsmath}

\def\NP#1#2{ Nucl.Phys. B#1 (#2)} 
\def\PL#1#2{ Phys.Lett. B#1 (#2)}
 
\def\MPL#1#2{ Mod.Phys.Lett.A#1 (#2)}

\def\HP#1#2{ JHEP #1 (#2)} 

\def\ap{ \alpha^{\prime}} 
\def\ai{\pi\alpha^{\prime}}
\def\pd{\partial}
\def\sa{\sigma}
\newcommand{\ep}{\text e}
\newcommand{\oh}{\frac{1}{2}}
\title{On Open String $\sigma$-Model and Noncommutative Gauge Fields}
\author{Oleg Andreev\thanks{e-mail:  andreev@physik.hu-berlin.de}
\thanks{Permanent address: Landau Institute, Moscow, Russia}
\hspace{.15cm} and Harald Dorn\thanks{e-mail: dorn@physik.hu-berlin.de}
\\ \\
Humboldt--Universit\"at zu Berlin, Institut f\"ur Physik\\
Invalidenstra\ss e 110, D-10115 Berlin, Germany}

\date{}
\begin{document} 
 
\maketitle 
\begin{abstract} 
We consider the ordinary and noncommutative Dirac-Born-Infeld theories within 
the open string $\sigma$-model. First, we propose a renormalization scheme, hybrid 
point splitting regularization, that leads directly to the Seiberg-Witten description including
the two-form $\Phi$. We also show how such a form appears within the standard 
renormalization scheme just by some freedom in changing variables. 
Second, we propose a Wilson factor which has the noncommutative gauge invariance on 
the classical level and then compute the $\sigma$-model partition function 
within one of the known renormalization scheme that preserves the noncommutative gauge 
invariance. As a result, we find the noncommutative Yang-Mills action.
\end{abstract}

\vspace{-12cm}
\begin{flushright}
hep-th/9912070       \\
HU Berlin-EP-99/64
\end{flushright}
\vspace{11cm}

\vspace{-.75cm}
\section{ Introduction} 
\renewcommand{\theequation}{1.\arabic{equation}}
\setcounter{equation}{0}
The open string $\sigma$-model was a hot topic in the eighties  as it provided a basic tool to derive 
the low energy effective action (Dirac-Born-Infeld action) that is nonperturbative in $\ap$ 
\cite{FT,C,P} \footnote{For a recent review of this issue see \cite{T} and references therein.}. Later 
it was also realized that it is useful for D-brane physics \cite{P1}. Recently, it has been pointed out by 
Seiberg and Witten that a special renormalization scheme, a point splitting regularization, results in a 
rather peculiar situation where the space-time (brane) coordinates do not commute (see \cite{SW} and 
refs. therein). The purpose of this paper is to further develop ideas about 
the appearance of noncommutative geometry by the open string $\sigma$-models.

Our conventions and some features of the quantization of open strings ending on D-branes 
that are relevant to our discussion are the following:

1. The $\sigma$-model action is given by 
\begin{equation}\label{ac}
S=\frac{1}{4\ai }\int_{\text{D}}d^2z\,\,
g_{ij}\pd_aX^i\pd^aX^j-i
\int_{\pd{\text D}}d\tau \,\Bigl(
\oh B_{ji}\dot X^iX^j+A_i(X)\dot X^i\Bigr)\,\,\,\,+\varphi
\quad,
\end{equation}
where D means the string world-sheet (disk or half plane) whose boundary is 
$\pd\text D$. $g_{ij},\,\,B_{ij},\,\,\varphi$ are the constant metric, antisymmetric tensor and dilaton fields, 
respectively. We also included in \eqref{ac} the Abelian background gauge field $A_i(X)$. $X^i$ map 
the world-sheet to the brane and $i,\,j=1,\dots , p+1$. The world-sheet indices are denoted by $a,b$.

At this point, let us make a couple of remarks. First, we keep the explicit dependence on the constant 
dilaton field because it plays an important role in what follows. Second, for a constant $B$ field 
$B_{ij}\epsilon^{ab}\pd_aX^i\pd_bX^j$ is a total derivative, so we write this term as a boundary 
interaction.

2. A natural object to compute within the path integral is the $\sa$-model partition function
\begin{equation}\label{pf}
Z[\varphi,g,B,A,\ap ]=\int {\cal D}X\,\ep^{-S}
\quad,
\end{equation}
that results in the open string low energy effective action (Born-Infeld action) as well as 
the D-brane action (Dirac-Born-Infeld action) (for a review and refs., see, \cite{T, P1}). 

To compute the partition function \eqref{pf} one first integrates over the internal 
points of the disc to reduce the integral to the boundary 
and next splits the integration variable $X^i$ in the constant and the non-constant part as 
$X^i(\tau)=x^i+\xi^i(\tau)$. As a result, one gets the infinite set of the vertices 
$F,\,\pd F,\pd^2F$ etc \footnote{For the sake of simplicity, we use the matrix notations here 
and below.}. Next the perturbation theory in $\ap$ is used to compute the leading 
terms of $Z$. The derivative-independent part of the partition 
function proves to be the Dirac-Born-Infeld action namely,
\begin{equation}\label{dbi}
Z[\varphi,g,B,A,\ap ] =S_{\text{DBI}}+O(\pd F)
=\ep^{-\varphi}\int [dx] \,\sqrt{\det (g+2\ai (F+B))}+O(\pd F)
\quad,
\end{equation}
where $F_{ij}=\pd_iA_j-\pd_jA_i$ and $[dx]=\frac{d^{p+1}x}{(2\ai )^{p+1}}$. The above result 
is based on the  use of one of the standard renormalization schemes like Pauli-Villars or $\zeta$-function 
which was originally used by Fradkin and Tseytlin \cite{FT}.

It is natural to ask, whether different renormalization schemes used to compute the path integral 
lead to the same answer. It was understood in the eighties that the ambiguity in the structure of 
$Z$ related to the choice of  a renormalization scheme should be a particular case of the field 
redefinition ambiguity present in the effective action reconstructed from the S-matrix. In 
other words, only structures in $Z$ that are invariant under the field redefinition are relevant.

3. Seiberg and Witten pointed out that in the framework of the point splitting regularization scheme
the leading terms of the perturbation theory are summed into the noncommutative version of the 
Dirac-Born-Infeld action namely,
\begin{equation}\label{ndbi}
\hat Z[\hat\varphi,G,\theta,\hat A,\ap ] 
=\hat S_{\text{DBI}}+\dots
=\ep^{-\hat\varphi}\int [dx]
\sqrt{\det (G+2\ai\hat F)}+\dots
\quad,
\end{equation}
where $\hat F_{ij}=\pd_i\hat  A_j-\pd_j\hat  A_i-i\hat A_i\ast\hat A_j+
i\hat A_j\ast\hat A_i$. Here the $\ast$-product is defined by 
\begin{equation}\label{star}
f(x)\ast g(x)=\ep^{\frac{i}{2}\theta^{ij}
\frac{\pd}{\pd y^i}\frac{\pd}{\pd z^j}}f(x+y)g(x+z)\vert_{y=z=0}
\quad.
\end{equation}
It is well-known that such a product is noncommutative but associative.

As we mentioned above, the different renormalization schemes are equivalent, so using a 
change of variables one has to get 
\begin{equation}\label{eq}
\hat S_{\text{DBI}}=S_{\text{DBI}}+\text{total derivative terms}+O (\pd F)
\quad.
\end{equation}
In the case of interest the change of variables found by Seiberg and Witten is given by \cite{SW}
\begin{equation}\label{nv}
\begin{split}
G&=(g-2\ai B)g^{-1}(g+2\ai B)
\quad\quad\,\,,\, \quad
\theta=-(2\ai )^2(g+2\ai B)^{-1}B(g-2\ai B)^{-1}
\quad, \\
\hat\varphi&=\varphi+\oh\ln\det \Bigl(G(g+2\ai B)^{-1}\Bigr)
\quad,\quad
\hat F=F+F\theta F-(A\theta\pd)F+O(\theta^2)
\quad.
\end{split}
\end{equation}
Note that the last relation is simplified in the case of a constant field $\hat F$. Explicitly,
\begin{equation}\label{F}
\hat F=(1+F\theta)^{-1}F
\quad.
\end{equation}

Moreover, it was also conjectured in \cite{SW} that the use of a suitable regularization that 
interpolates between the both mentioned above results in \footnote{It should be noted that 
though analogous formulae first appeared in the context of NCSYM \cite{Pio} their 
appearance within the $\sigma$-model approach is due to Seiberg and Witten.}
\begin{equation}\label{fdbi}
\mathbf{Z}[\boldsymbol{\varphi},\mathbf{G},\boldsymbol{\theta},\mathbf{A},\ap ] 
=\mathbf{S}_{\text{DBI}}+\dots
=\ep^{-\boldsymbol{\varphi}}\int [dx]
\sqrt{\det (\mathbf{G}+2\ai(\mathbf{F}+\boldsymbol{\Phi}))}+\dots
\quad,
\end{equation}
where $\boldsymbol{\Phi}$ is some two-form. In this case the relations \eqref{nv} are modified to 
\begin{equation}\label{fnv}
(\mathbf{G}+2\ai\boldsymbol{\Phi})^{-1}=
-\frac{1}{2\ai }\boldsymbol{\theta}+
(g+2\ai B)^{-1}
\quad,\quad
\boldsymbol{\varphi}=\varphi+\oh\ln\det\Bigl(\frac{\mathbf{G}+2\ai\boldsymbol{\Phi}}
{g+2\ai B}\Bigr)
\quad.
\end{equation}

\section{ Seiberg-Witten conjecture via $\sigma$-model} 
\renewcommand{\theequation}{2.\arabic{equation}}
\setcounter{equation}{0}
The aim of this section is to show how the Seiberg-Witten conjecture \eqref{fdbi} may be simply 
derived within the $\sigma$-model approach. First, we start from the point splitting 
renormalization scheme. Then we propose how to modify it to get the desired result. 
Second, we start with the $\zeta$-function renormalization scheme and use the change of 
variables  ($\sigma$-model couplings) to get \eqref{fdbi}. In the both cases $\Phi$ naturally 
appears due to freedom in choosing new variables. 

\subsection{Hybrid point splitting regularization}

Following the ideas sketched in the introduction, we split the integration variable $X^i$  and moreover 
include the $B\dot\xi\xi$ term into the kinetic term that is usual for the problem at hand \cite{C,SW}.  
So we have for the partition function \eqref{pf}
\begin{equation}\label{pf-ps}
\begin{split}
&Z[\varphi,g,B, A,\ap ]
=\ep^{-\varphi}\int [dx]\,\,
\langle\,\ep^{-S_{int}}\,\rangle
\quad,\quad
\langle\,\dots\,\rangle=\int{\cal D\xi}\,\ep^{-S_0}\dots
\quad, \\
&S_0=\oh\int_{\pd{\text D}}d\tau d\tau^{\prime}\,
\xi^{\text{\tiny T}}\hat N^{-1}\xi
\quad,\quad
S_{int}=-i\int_{\pd{\text D}}d\tau \, A(x+\xi)\dot \xi
\quad,
\end{split}
\end{equation}
where $\hat N=-\ap G^{-1}\ln(\tau -\tau^{\prime})^2+\frac{i}{2}\theta
\epsilon (\tau-\tau^{\prime})$. The matrices $G,\,\theta$ are given by Eq. \eqref{nv}.

Expanding the interaction part in $A$ and doing the Fourier transform, the partition function is 
\begin{equation}\label{pf-ps1}
Z[\varphi,g,B, A,\ap ]=
\ep^{-\varphi}\int [dx]
\sum_{n=0}^{\infty}\frac{i^n}{n!}\idotsint d\tau_n \idotsint d^{p+1}k_n \,\,
\ep^{i\kappa x}
A(k_1)\dots A(k_n)\,\,\langle\,\prod_{j=1}^n\dot\xi
\ep^{ik_j\xi (\tau_j)}\,
\rangle_{G,\,\theta}
\,\,,
\end{equation}
where $\kappa=k_1+\dots +k_n$.

From now let us stick to the point splitting regularization.  We are aware of the problem to define what is 
meant by the point splitted path integral for the partition function. Instead of giving a general answer, we 
define for our purpose a hybrid point splitting regularization. As used in \cite{SW} the point splitting 
regularization is not defined per se, but only in connection with a partial summation of perturbation theory 
that includes $B\dot\xi\xi$ term into the kinetic term. This results in an additive part $\frac{i}{2}\theta
\epsilon(\tau-\tau^{\prime})$ in the corresponding propagator. Everywhere the propagator ends on 
vertices which contain a $\tau$-derivative, part of the regularization is to drop the $\pd\epsilon$ term. 
We use this rule throughout, also for possible operator expressions inside functional determinants. But we 
insist that the determinants for the remaining expressions are treated within the standard renormalization 
scheme like Pauli-Villars or $\zeta$-function. As a consequence we get a generalization of the 
corresponding formula in \cite{SW}
\begin{equation}\label{path}
\langle\,\prod_{j=1}^n\dot\xi
\ep^{ik_j\xi (\tau_j)}\,\rangle_{G,\,\theta}=J\,
\ep^{-\frac{i}{2}\sum _{i>j}k_i\boldsymbol{\theta}k_j
\epsilon (\tau_i-\tau_j)}
\,\langle\,\prod_{j=1}^n\dot\xi
\ep^{ik_j\xi (\tau_j)}\,
\rangle_{G,\,\theta_0}
\quad,
\end{equation}
where $\boldsymbol{\theta}=\theta-\theta_0$. As to $J$, it is the quotient of functional determinants. 
We have to care of it because it is relevant for what follows. It is evident that there exists freedom in 
choosing the new parameter $\theta_0$. It turns out that exactly such freedom is responsible for 
the two-form $\Phi$. Let us go on to see that this is indeed the case. Undoing the Fourier transform 
we arrive at \footnote{We change our notations due to specializing the renormalization scheme. So, 
$\mathbf{A}$ denotes the gauge variable in this scheme etc.}
\begin{equation}\label{pf-ps2}
\mathbf{Z}[\varphi,\mathbf{G},\boldsymbol{\theta},\mathbf{A},\ap ]= J \ep^{-\varphi}\int [dx]
\sum_{n=0}^{\infty}
\idotsint d\tau_n\,  
\ep^{\frac{i}{2}\sum_{i>j}\pd^i\boldsymbol{\theta}\pd^j
\epsilon (\tau_i-\tau_j)} 
\langle\, \prod_{j=1}^n
\mathbf{A}(x_j+\xi_j)\dot \xi (\tau_j)\vert_{x_1=\dots x_n=x}
\,\rangle_{G,\,\theta_0}
\,\,,
\end{equation}
where $\pd^i=\frac{\pd}{\pd x_i}$.

An important point we should stress here is that the above redefinition of $\theta$ automatically 
leads to a proper redefinition of the kinetic term. This time we have
\begin{equation}\label{k}
\mathbf{S}_0=\oh\int_{\pd{\text D}}d\tau d\tau^{\prime}\,
\xi^{\text{\tiny T}} \mathbf{G}N^{-1}\xi
-\frac{i}{2}\int_{\pd{\text D}}d\tau \,\xi^{\text{\tiny T}}\boldsymbol{\Phi}\dot\xi
\quad,\quad
N=-\ap\ln(\tau-\tau^{\prime})^2
\quad,
\end{equation}
with the new matrices for the metric and antisymmetric field, given by the corresponding inversion of the 
relations \eqref{nv}
\begin{equation}\label{met-b}
\mathbf{G}=\Bigl(G^{-1}-\frac{1}{(2\ai )^2}\theta_0G\theta_0\Bigr)^{-1}
\quad,\quad
\boldsymbol{\Phi}=-\frac{1}{(2\ai )^2}\mathbf{G}\theta_0G
\quad.
\end{equation}
A simple algebra shows that the such defined $\boldsymbol{\theta},\,\mathbf{G},\,\boldsymbol{\Phi}$ 
obey the Seiberg-Witten relation \eqref{fnv}. 

Since our regularization prescription allows us to treat the functional determinants along the lines 
of \cite{FT} we find for $J$ 
\begin{equation}
J=
\Biggl(\frac{\text{Det}\hat N_{G,\theta}}{\text{Det}\hat N_{G,\theta_0}}
\Biggr)^{\oh }
=
\Biggl(\frac{\det (g+2\ai B)}{\det (\mathbf{G}+2\ai\boldsymbol{\Phi})}\Biggr)^{\oh }
\quad.
\end{equation} 
The latter may be absorbed by the corresponding redefinition of the dilatonic field (string coupling constant). 
Thus we get the last relation in \eqref{fnv}.

Let us now define the Wilson factor as 
\begin{equation}\label{Wl}
P\ast \exp \Bigl(i\int\,\mathbf{A}(x+\xi)\dot \xi\Bigr)=\sum_{n=0}^{\infty}\frac{i^n}{n!}
\idotsint d\tau_n\,  
\ep^{\frac{i}{2}\sum_{i>j}\pd^i\boldsymbol{\theta}\pd^j
\epsilon (\tau_n-\tau_m)} 
\prod_{j=1}^n
\mathbf{A}(x_j+\xi_j)\dot \xi (\tau_j)\vert_{x_1=\dots x_n=x}
\quad.
\end{equation}
It is clear that this factor is equal to the usual path ordering $P$ applied to the exponential in 
the sense of the $\ast$-product.

Finally, the partition function becomes 
\begin{equation}\label{pf-ps3}
\mathbf{Z}[\boldsymbol{\varphi},\mathbf{G},\boldsymbol{\theta},\mathbf{A},\ap ]
=\,\ep^{-\boldsymbol{\varphi}}
\int [dx]\int{\cal D}\xi\,
\ep^{-\mathbf{S}_0}\,P\ast \exp \Bigl(i\int_{\pd{\text D}}d\tau\,\mathbf{A}(x+\xi)\dot\xi\Bigr)
\quad.
\end{equation}

Let us now specialise the gauge field to 
\begin{equation}\label{cf}
\mathbf{A}_i(X)=\frac{1}{2}\mathbf{f}_{ji}X^j
\quad,
\end{equation}
with a constant matrix $\mathbf{f}$. Then the related noncommutative field strength is simply given by
$\mathbf {F}=\mathbf{f}-\mathbf{f}\boldsymbol{\theta}\mathbf{f}$. 

For the gauge field \eqref{cf} the Wilson factor \eqref{Wl} turns out to be
\begin{equation}\label{cfW}
P\ast \exp \Bigl(\frac{i}{2}\int_{\pd{\text D}}d\tau\,(x+\xi)^{\text{\tiny T}}\mathbf{f}\dot\xi\Bigr)=
\exp \Bigl(\frac{i}{2}\int_{\pd{\text D}}d\tau\,(x+\xi)^{\text{\tiny T}}\mathbf{F}\dot\xi\Bigr)
\quad.
\end{equation}
Now we just have the functional integral treated in \cite{FT} and find
\begin{equation}\label{fin}
\mathbf{Z}[\boldsymbol{\varphi},\mathbf{G},\boldsymbol{\theta},\mathbf{A},\ap ]=
\mathbf{S}_{\text{DBI}}=
\,\ep^{-\boldsymbol{\varphi}}
\int [dx]
\sqrt{\det (\mathbf{G}+2\ai(\mathbf{F}+\boldsymbol{\Phi}))}
\quad.
\end{equation}
With this formula we  have finished an explicit calculation proving that the 
partition function for a constant field strength within the hybrid point 
splitting renormalization scheme
is given by the Dirac-Born-Infeld action referring to the noncommutative gauge field.
This has been argued for in \cite{SW} by referring to the analogy of the
involved structures compared to the commutative case.  

\subsection{$\zeta$-function regularization}
As we have mentioned in the introduction, the different renormalization schemes used to 
compute the path integral \eqref{pf} lead to the same result after a proper change of the variables 
($\sigma$-model couplings) is done. In fact, such a change of the couplings is nothing but the 
resummation of the perturbation theory in $\ap$. After this is understood, it immediately comes to 
mind to realize a resummation by changing the path integral variables. A possible way to do this is 
to take a new variable as a function of $\ap$. Let us now show how it works. Specialising to the 
$\zeta$-function renormalization scheme \cite{FT}, we define the path integral measure in the same 
way as it was done in \cite{AMT}. So we have
\begin{equation}\label{pt}
\begin{split}
&Z[\varphi,g,B,A,\ap ]=\ep^{-\varphi}\int [dx]\sqrt{\det g}\,\,
\langle\,\ep^{-S_{int}}\,\rangle
\quad,\quad
\langle\,\dots\,\rangle=\int{\cal D\xi}\,\ep^{-S_0}\dots
\quad, \\
&S_0=\oh\int_{\pd{\text D}}d\tau d\tau^{\prime}\,
\xi^{\text{\tiny T}}gN^{-1}\xi
\quad,\quad
S_{int}=-i\ai\int_{\pd{\text D}}d\tau \, \xi^{\text{\tiny T}}(B+F)\dot\xi+O(\pd F)
\quad,
\end{split}
\end{equation}
Above we have rescaled $X^i$ as $X^i\rightarrow\sqrt{2\ai }X^i$. $N$ is the boundary value of the 
Neumann function. 

Now let us change the variable $\xi^i$ as $\xi^i=(g^{-1}\Lambda\boldsymbol{\xi})^i\,$ 
\footnote{In fact, it is nothing but $GL(p+1)$ transforms whose matrices depend on $\ap$.}. In fact, the 
measure is defined in such a way that the effect is only due to the action $S$. The latter becomes
\begin{equation}\label{na}
S_0=\oh\int_{\pd{\text D}}d\tau d\tau^{\prime}\,
\boldsymbol{\xi}^{\text{\tiny T}}\mathbf{G}N^{-1}\boldsymbol{\xi}
\quad,\quad
S_{int}=-i\ai\int_{\pd{\text D}}d\tau \, \boldsymbol{\xi}^{\text{\tiny T}}\bigl(
-\frac{1}{(2\ai )^2}\mathbf{G}\boldsymbol{\theta}\mathbf{G}+\Lambda^{\text{\tiny T}}g^{-1}F
g^{-1}\Lambda\bigr)\boldsymbol{\xi}+O(\pd F)
\,\,,
\end{equation}
where $\mathbf{G}=\Lambda^{\text{\tiny T}}g^{-1}\Lambda,\,
\boldsymbol{\theta}=-(2\ai )^2\Lambda^{-1}B\Lambda^{\text{\tiny T}-1}$.

It is evident that as far as $\Lambda$ depends on $\ap$ we automatically get a resummation of the 
perturbation theory 
because the vertices ($\sigma$-model couplings) are redefined \footnote{Note that 
$\Lambda=g+2\ai B$ recovers the Seiberg-Witten relations \eqref{nv}.}. It is also clear which role 
the dilaton 
field $\varphi$ has to play. It is responsible for the cancellation of the poles in $\ap$ within the 
perturbation theory i.e., the $\ap$-dependence of $Z$ always looks like 
$Z\sim (\ap )^{-\frac{p+1}{2}}\Bigl(1+O(\ap )\Bigr)$. Note that $\Lambda$ is an arbitrary 
function in $g,\,B,\,\ap $. Moreover $B$ plays a key role 
as it is responsible for $\ap$-dependence of $\Lambda$ because $g$ is dimensionless. One can 
also consider $\Lambda$ as a function 
in $\mathbf{G},\,\theta,\,\ap $. In this case the change of the variables is given by 
$\xi^i=(\Lambda^{\text{\tiny T}-1}\mathbf{G}\boldsymbol{\xi})^i$.

In general, it is not clear how to exactly compute the partition function. The best what we can do is to 
find the leading terms within the perturbation expansion as it was done in \cite{FT,AT,T2}. For our 
purposes  the one-loop approximation is also sufficient. So the problem is reduced to finding the 
corresponding determinant. It is straightforward to write down a solution of the problem \cite{FT}
\begin{equation}\label{det}
S_{\text{DBI}}=\ep^{-\varphi}\int [dx]\sqrt{\det g}\,\,
\sqrt{\det (1-\frac{1}{2\ai }\boldsymbol{\theta} \mathbf{G}+
2\ai\Lambda^{-1}F\Lambda^{\text{\tiny T}-1}\mathbf{G}\,)}
\quad.
\end{equation}
In doing transformations with the determinant the important thing one should keep in mind is that 
the resummation of the perturbation theory assumes that the partition function depends on $\ap$ as 
$Z\sim (\ap )^{-\frac{p+1}{2}}\Bigl(1+O(\ap )\Bigr)$. As to the determinant, we postulate that it 
is given by $\sqrt{\det (A+2\ai B)}$ with some matrices $A$ 
and $B$. So it is clear that we should get rid of the $\frac{1}{\ap}$-term. The latter argument assumes 
\begin{equation}\label{det1}
S_{\text{DBI}}=\ep^{-\varphi}\sqrt{\det (g+2\ai B)\,}
\int [dx]\,\,
\sqrt{\det (1+(1-\frac{1}{2\ai }\boldsymbol{\theta}\mathbf{G})^{-1}
2\ai\Lambda^{-1}F\Lambda^{\text{\tiny T}-1}\mathbf{G}\,)}
\quad.
\end{equation}
Above we have used that $\det(g-\frac{1}{2\ai }g\boldsymbol{\theta}\mathbf{ G})=\det (g+2\ai B)$.

Moreover we bring $\Lambda^{-1}$ to a form 
\begin{equation}\label{del}
\Lambda^{-1}=(\mathbf{G}^{-1}-\frac{1}{2\ai }\boldsymbol{\theta} )\Delta
\quad,
\end{equation}
where $\Delta$ is a new matrix that depends on $\mathbf{G},\,\boldsymbol{\theta},\,\ap $. This 
results in 
\begin{equation}\label{det2}
S_{\text{DBI}}=\ep^{-\varphi}\sqrt{\det (g+2\ai B)\,}
\int [dx]\,\,
\sqrt{\det (1+2\ai F\Delta^{\text{\tiny T}}(\mathbf{G}^{-1}+\frac{1}{2\ai }\boldsymbol{\theta})
\Delta)\,}
\quad.
\end{equation}
Our postulate for the determinant yields the following $\ap$-dependence of 
$\Delta^{\text{\tiny T}}(\mathbf{G}^{-1}+\frac{1}{2\ai }\boldsymbol{\theta})\Delta$
\begin{equation}\label{del1}
\Delta^{\text{\tiny T}}(\mathbf{G}^{-1}+\frac{1}{2\ai }\boldsymbol{\theta})\Delta
=\frac{1}{2\ai }\Gamma+\Sigma(\ap )
\quad,
\end{equation}
where $\Sigma(\ap )$ is regular in the limit $\ap\rightarrow 0$.

Substituting \eqref{del1} into \eqref{det2} we find
\begin{equation}\label{det3}
S_{\text{DBI}}=\ep^{-\boldsymbol{\varphi}}
\int [dx]\,\,
\sqrt{\det(1+F\Gamma )\,}
\sqrt{\det (\Sigma^{-1}+2\ai (1+F\Gamma)^{-1}F\,)}
\quad.
\end{equation}
Above we have also defined a new dilatonic field as 
\begin{equation}\label{dilaton}
\boldsymbol{\varphi}=\varphi+\oh \ln\frac{\det \Sigma^{-1}}{\det(g+2\ai B)}
\quad.
\end{equation}

Furthermore due to our postulate for the determinant we can 
represent $\Sigma^{-1}(\ap )$ as $\Sigma^{-1}(\ap )=\Pi+2\ai\Omega$ with some 
$\ap$-independent matrices $\Pi$ and $\Omega$. So the Eq. \eqref{det3} is rewritten in the 
following form
\begin{equation}\label{det4}
S_{\text{DBI}}=\ep^{-\boldsymbol{\varphi}}
\int [dx]\,\,
\sqrt{\det(1+F\Gamma)\,}
\sqrt{\det (\Pi+2\ai (\Omega+(1+F\Gamma)^{-1}F)\,)}
\quad.
\end{equation}

Let us now see what the matrices $\Sigma,\Pi,\Omega$ are. To do so, it is worth to remark that 
$\Lambda_{ij}=g_{ij}$ is a trivial 
transform in a sense that it does not lead to any resummation of the perturbation theory. So we can 
consider it as a unity in a space of all transforms. It is reasonable to normalize transforms with 
respect of 
this unity by requiring that $\Delta$ has the following asymptotic behaviour $\Delta_i^j=\delta_i^j
+O(\ap )$ as $\ap\rightarrow 0$. On the other hand, it also seems reasonable because $\delta_i^j$ is 
the only $\ap$-independent tensor with such an index structure we have. Then it immediately 
follows 
from \eqref{del1} that $\Gamma=\boldsymbol{\theta}$. This allows to define a new 
field strength $\mathbf{F}$ as 
$\mathbf{F}=(1+F\boldsymbol{\theta})^{-1}F$ that of course coincides with the Seiberg-Witten 
definition 
\eqref{F}. 
However, the point is that we have not used the fact that two theories should be in the same class of 
equivalence as gauge theories. As a result, the Dirac-Born-Infeld action is rewritten as 
\begin{equation}\label{det5}
S_{\text{DBI}}=\ep^{-\boldsymbol{\varphi}}
\int [dx]\,\,
\sqrt{\det(1+F\boldsymbol{\theta})\,}
\sqrt{\det (\Pi+2\ai (\Omega+\mathbf{F})\,)}
\quad.
\end{equation}
Furthermore it is clear from the general covariance of the partition function that $\Pi$ must be 
treated as a metric $\mathbf{G}$. As to $\Omega$, it coincides with $\Phi$ 
from Eq.\eqref{fdbi}. Of course, \eqref{del}, \eqref{del1}, \eqref{dilaton} are equivalent to the 
relations \eqref{fnv}.

Finally, we get
\begin{equation}\label{det6}
S_{\text{DBI}}=\ep^{-\boldsymbol{\varphi}}
\int [dx]\,\,
\sqrt{\det(1+F\boldsymbol{\theta})\,}
\sqrt{\det (\mathbf{G}+2\ai (\Phi+\mathbf{F})\,)}
\quad.
\end{equation}

What we actually need is only the last factor of the integrand to get the equivalence \eqref{eq}. To fix 
this problem, one subtle point 
we should remind is that different formulations coincide up to derivative terms. It also assumes 
that we can integrate by parts that of course makes no sense for constant field strengths. In fact, 
we have to consider slowly varying field strengths. As a consequence, the relation between $F$ and 
$\mathbf{F}$ becomes more involved \cite{SW}. In the leading order of $\boldsymbol{\theta}$ 
it is given by
\begin{equation}\label{dF}
\mathbf{F}=F-F\boldsymbol{\theta}F-(A\boldsymbol{\theta}\pd)F+O(\boldsymbol{\theta}^2)
=\mathbf{F}_0-(A\boldsymbol{\theta}\pd)F+O(\boldsymbol{\theta}^2)
\quad.
\end{equation}
It is easy to see that in this order the actions \eqref{fdbi} and \eqref{det6} coincide namely,
\begin{equation*}
\int [dx]\,\biggl(\sqrt{\det(1+F\boldsymbol{\theta})\,}
\sqrt{\det (G+2\ai (\Phi+\mathbf{F}_0 )\,)}
-\sqrt{\det (G+2\ai (\Phi+\mathbf{F}_0-(A\boldsymbol{\theta}\pd)F)\,)}\biggr)
=0
\quad.
\end{equation*}
 Unfortunately calculations become more and more involved as far as we consider higher order terms.

To conclude, let us comment on the background independence of the action 
$\mathbf{S}_{\text{DBI}}$. Since the action $S_{\text{DBI}}$ is explicitly invariant the 
expression \eqref{det6} we find in above is also invariant. Next what we drop is total derivative terms, 
so the action $\mathbf{S}_{\text{DBI}}$ is invariant at least in the leading order in 
$\boldsymbol{\theta}$ modulo total derivatives.


\section{ Alternative way} 
\renewcommand{\theequation}{3.\arabic{equation}}
\setcounter{equation}{0}
In fact, there are two ways in 
getting the noncommutative Yang-Mills theory with the $\ast$-product structure 
within the $\sigma$-model. The first one {\it a la} Seiberg-Witten is to start from the 
Wilson factor that has the ordinary gauge invariance on the classical level and then get the 
noncommutative gauge invariance on the quantum level just by using a proper renormalization 
scheme. The second way we propose, based on our experience with the hybrid renormalization scheme, 
is to start from the Wilson factor that has the noncommutative 
gauge invariance on the classical level and use a regularization that maintains it on the quantum 
level \footnote{ Some similar motivations are provided by the quantization of open strings in a constant 
$B$ field background where the deformation parameter $\theta$ is explicitly related with zero modes 
(see, e.g., \cite{Fa} and refs. therein). }. So our aim in this section is to show how to realize this proposal. 

Let us define the Wilson factor as 
\begin{equation}\label{nW}
W[C]=P\ast \exp \Bigl(i\int_Cd\tau\,A\dot X\Bigr)
=\sum_{n=0}^{\infty}i^n
\idotsint d\tau_n\, H(\tau_{12})\dots H(\tau_{n-1n})A\dot X(\tau_1)\ast\dots\ast
\,A\dot X(\tau_n)
\quad,
\end{equation}
where the $\ast$-product is defined with respect to a translational mode of $X$. Such a factor 
coincides with the one defined in \eqref{Wl}. This is clear  just 
by substituting the expansion of unity $1=H(\tau_{12})\dots H(\tau_{n-1n})+(\text{all perms.})$ 
\footnote{$H$ means the Heaviside step function.} and 
changing the variables $\tau$ in such a way to get the ordering $\tau_1>\tau_2>\dots>\tau_n$. 
Note that the definition is nothing but a slightly modified version of the non-Abelian Wilson 
factor. Of course, it is simply to fit the non-Abelian case into the above definition just by allowing the 
gauge field to be a $N\times N$ hermitian matrix and taking a trace. Explicitly, 
\begin{equation}\label{naW}
W[C]=TrP\ast \exp \Bigl(i\int_Cd\tau\,A\dot X\Bigr)
\quad.
\end{equation}

An important point we should stress here is that the Wilson factor as it is defined in \eqref{nW} 
or \eqref{naW} is {\it almost} invariant \footnote{It becomes gauge invariant after the integration over the 
translational mode of $X$.} under the gauge transformation of 
the noncommutative Yang-Mills theory 
namely, $\delta_{\lambda}A_i=\pd_i\lambda-iA_i\ast\lambda +i\lambda\ast A_i$. The latter is 
clear from an analogy with the non-Abelian Yang-Mills theory where the gauge invariant expression 
is given by a trace of $P\exp\int A{\dot X}$. Of course, this is easy to see by directly doing the 
gauge transforms. Fortunately for us, what saves the day is that we are interested in the 
partition function. Indeed, as we saw splitting the integration variable $X^i$ automatically provides 
the integral over the zero mode (translational mode) $x^i$. This is exactly what we need because 
in the noncommutative case the integral does the same job as the trace in the non-Abelian case, i.e. it 
provides the cyclic 
property $\int d^{p+1}x\,f(x)\ast g(x)=\int d^{p+1}x\,g(x)\ast f(x)$ that is crucial for the gauge 
invariance. Thus the partition function is gauge invariant.  

To be more precise, the partition function is given by 
\begin{equation}\label{z}
Z[A]=\int{\cal D}X\,\ep^{-S_0}\,P\ast \exp \Bigl(i\int_Cd\tau\,A_i\dot X^i\Bigr)
\quad,\quad
S_0=\frac{1}{4\ai }\int_{\text{D}}d^2z\,\,G_{ij}\pd_aX^i\pd^aX^j
\quad.
\end{equation}

To see that this partition function indeed leads to the noncommutative Yang-Mills theory 
it is instructive to compute its asymptotic behaviour as $\ap\rightarrow 0$. The main point is that 
we have to be careful to preserve the noncommutative gauge invariance. It turns out that the 
renormalization scheme based on the $\zeta$-function or some of its modification does this job \cite{FT,AT}. 
The computations are analogous to the ordinary non-Abelian case \cite{T2}. As a result, we find
\begin{equation}\label{nYM}
Z[A]=\int [dx]\sqrt{\det G\,}
\Bigl(1+\frac{1}{4}(2\ai )^2F_{ij}\ast F_{kl}G^{ik}G^{jl}+
O(\ap{}^{\frac{5}{2}}) \Bigr)
\quad,
\end{equation}
with $F_{ij}=\pd_iA_j-\pd_jA_i-iA_i\ast A_j +iA_j\ast A_i$. 

It is straightforward to generalize these computations for the non-Abelian case. The only new thing 
that appears is $Tr$. 

Finally, let us briefly show how to incorporate SUSY within the above formalism. For simplicity, let us 
consider the NSR formalism. In other words, we add a set of the fermionic fields $\psi^i$ whose 
metric also is $G$. Following \cite{AT}, it is simply to suggest what the Wilson factor should be. Moreover, 
the formalism developed in this paper allows to use the super-space notations. 
Thus, the Wilson factor is given by
\begin{equation}\label{snW}
\mathbf{W}[C]=\mathbf{P}\ast \exp \Bigl(i\int_Cd\boldsymbol{\tau}\,A D\mathbf{X}\Bigr)
=\sum_{n=0}^{\infty}i^n
\int .\,.\,.\int
d\boldsymbol{\tau}_n\mathbf{H}(\boldsymbol{\tau}_{12})\,.\,.\,.\,
\mathbf{H}(\boldsymbol{\tau}_{n-1n})\,
AD\mathbf{X}(\boldsymbol{\tau}_1)
\ast\dots\ast
AD\mathbf{X}(\boldsymbol{\tau}_n)
\,,
\end{equation}
where we use the super-space notations namely, $d\boldsymbol{\tau}=d\tau d\vartheta,\,
\mathbf{H}(\boldsymbol{\tau}_{nm})=H(\tau_{nm})+\vartheta_n\vartheta_m\delta(\tau_{nm}),\,
\mathbf{X}^i=X^i+\vartheta\psi^i,\,D=\vartheta\pd_{\tau}-\pd_{\vartheta}$. As in the bosonic case 
the noncommutative multiplication law is defined in terms of the translational modes of $X^i$.

It is a simple matter to check that the expression \eqref{snW} is {\it almost} 
gauge invariant as well as to compute the partition function in the leading order in $\ap$. The result is 
again given by the noncommutative Yang-Mills  action (see Eq.\eqref{nYM}). It is also straightforward 
to fit the non-Abelian case into the above definition just by doing in the same way way as we 
did in the bosonic case.


\section{ Concluding Comments} 
\renewcommand{\theequation}{4.\arabic{equation}}
\setcounter{equation}{0}
Motivated by our experience with the $\sigma$-model, we would like to propose 
an exactly gauge invariant version of the Wilson factor within the noncommutative Yang-Mills theory,
\begin{equation}\label{nW-loop}
{\cal W}[C]=\frac{1}{V}\int d^{p+1}x\sqrt{\det G\, }\,\, TrP\ast \exp \Bigl(i\int_Cd\tau\,A\dot X
\Bigr)
\quad,\quad \text{ with}\quad V=\int d^{p+1}x\sqrt{\det G\,}
\quad.
\end{equation}

Some discussions of the Wilson factors and Dirac-Born-Infeld action that have some interference with 
what we described in above are due to \cite{refs}.

\vspace{.25cm} {\bf Acknowledgements}

\vspace{.25cm} 
We would like to thank A. Tseytlin for comments and reading the manuscript.
The work of O.A. is supported in part by the Alexander von Humboldt Foundation, 
and by Russian Basic Research Foundation under Grant No. 9901-01169.
The work of H.D. is supported in part by DFG.



\end{document}